\documentclass{iaus}

\usepackage{graphicx}
\usepackage{natbib}
\usepackage{multicol}

\newcommand{\ARAA}{\textit{ARAA}}

\newcommand{\aj}{\textit{AJ}}                
\newcommand{\apj}{\textit{ApJ}}               
\newcommand{\apjl}{\textit{ApJL}}    
\newcommand{\apjs}{\textit{ApJS}}

\newcommand{\AnA}{\textit{A\&A}}               
\newcommand{\aap}{\textit{A\&A}}               
              
\newcommand{\AAS}{\textit{A\&AS}}

\newcommand{\mnras}{\textit{MNRAS}}

\newcommand{\PhysRevD}{\textit{Phys. Rev. D}}     
\newcommand{\PhysRevLett}{\textit{Phys. Rev. Lett.}}

\newcommand{\ssrv}{\textit{Space Sci. Revs}}

\title[S 262.~~Stellar populations challenges]{Challenges in Stellar
  Population Studies}

\author[Jarle Brinchmann]{Jarle Brinchmann$^1$}
\affiliation{$^1$Leiden Observatory, Leiden University,\\ P.O. Box
  9513, 2300 RA, Leiden, The Netherlands \\ email:
  \texttt{jarle@strw.leidenuniv.nl}}
\pubyear{2009}
\volume{262}
\pagerange{\ldots}
\jname{Stellar Populations: Planning for the Next Decade}
\editors{Gustavo Bruzual \& St{\'e}phane Charlot, eds.}

\def\urltilda{\kern -.05em\lower .7ex\hbox{\~{}}\kern -.1em}

\newcommand{\Msun}{\ensuremath{M_{\odot}}}

\begin{document}

\maketitle

\begin{abstract}
  The stellar populations of galaxies contain a wealth of detailed
  information. From the youngest, most massive stars, to almost
  invisible remnants, the history of star formation is encoded in the
  stars that make up a galaxy. Extracting some, or all, of this
  information has long been a goal of stellar population studies. This was achieved in the last couple of decades and  it is now a routine task, which 
   forms a crucial ingredient in much of observational galaxy
  evolution, from our Galaxy out to the most distant systems
  found.  In many of these domains we are now limited not by sample
  size, but by systematic uncertainties and this will increasingly be
  the case in the future.

  The aim of this review is to outline the challenges faced by stellar
  population studies in the coming decade within the context of
  upcoming observational facilities.  I will highlight the need to better understand the near-IR spectral range and outline the difficulties presented by less well understood phases of stellar evolution such as thermally pulsing AGB stars, horizontal branch stars and the very first stars. The influence of rotation and binarity on stellar population modelling is also briefly discussed.
\end{abstract}

\firstsection
\section{Introduction}
\label{sec:introduction}

The luminous output of normal galaxies is ultimately generated by stars in
various stages of stellar evolution --- in isolation a trite
observation, but this simple fact gives us the opportunity to extract a vast
amount of information from observations of galaxies through the
modelling of their stellar populations.

While the study of stellar populations in galaxies started with
Baade's (1944)\nocite{1944ApJ...100..137B} identification of two 
populations of stars in M32 and NGC 205, a 
rigourous study of the topic only commenced in the late 60's and early
70's \citep{Tinsley1968,1972A&A....20..361F,1973ApJ...179..427S} with
Tinsley's \textit{Fundamentals of Cosmic Physics} article
\citep{1980FCPh....5..287T} particularly influential.  Since that 
time, the number of articles discussing stellar populations has risen
rapidly so that today about 12\% of all articles in the major journals mention
stellar populations in their abstracts (Figure 1).

The majority of this growth has been made possible through the
development of simple models for the evolution of stellar populations
that have found widespread use in a wide range of astronomical
studies, from stellar clusters in the Milky Way to the most distant
galaxies in the Universe. I will repeatedly refer to stellar population models below, and use this term loosely to refer to any model that predict the observational properties of any ensemble of stars. These start from models of stellar evolution (e.g. Bertelli et al 1994; see contribution by Cassisi these proceedings). By applying empirical colour corrections (e.g. Lejeune et al 1997) they can be placed on an observational Hertzsprung-Russell diagram. For a given star formation history and initial mass function (IMF), this can be sampled, either to produce a Monte Carlo realisation of an observed colour-magnitude diagrams (see Tolstoy et al 2009), or by convolution to create integrated properties of a stellar population (e.g. Fioc \& Rocca-Volmerange 1997; Leitherer et al 1999; Vazdekis 1999; Bruzual \& Charlot 2003;  Maraston 2005; Kotulla et al 2009). Finally, these models are compared to observations using a number of techniques, in itself an interesting topic, but for this review the details of this step are not crucial.

In this contribution I will begin by briefly reviewing some of the achievements in extra-galactic astronomy made possible by our understanding of stellar populations, before I turn to the challenges for stellar populations studies in the coming decade. Given the format this will not be an exhaustive review, see the other contributions in these proceedings for more in-depth discussion of many of the topics.

\section{A series of successes}
\label{sec:series-successes}

\subsection{Resolved stellar populations --- extracting star formation histories}
\label{sec:resolv-stell-popul}

From the very beginning, the study of objects where individual stars can be
resolved has been central to stellar population modelling. The reason
for this is that one can construct a colour-magnitude diagram
(CMD) directly from the observations and this is sufficiently close to
the theoretical Hertzsprung-Russell diagram that insight into the
properties of the population of stars can be had fairly
straightforwardly.  

By comparing a theoretical CMD to the observational one, it is
possible to infer a number of properties for the stellar system being
studied. The first such study was arguably that by \citet{Maeder1974} but
the full power of the approach has only been realised in the last
couple of decades, with an early application of this technique to
infer the star formation history of Local Group dwarfs by
\citet{1991AJ....102..951T}.  

The field blossomed through the use of Hubble Space Telescope (HST) to
obtain very deep CMDs of dwarf galaxies throughout the Local Group
(see Tolstoy, Hill \& Tosi 2009 for a
review)\nocite{2009arXiv0904.4505T} and through the 
development of sophisticated algorithms for the analysis of CMDs
\citep[see][]{Gallart2005}. This wealth of data has provided
us with an impressive insight into the history of star formation in
small galaxies in the Local Group and it is now clear that they show
an enormous range in star formation histories. 


\begin{figure}[tbp]
\begin{center}
\includegraphics[width=\textwidth]{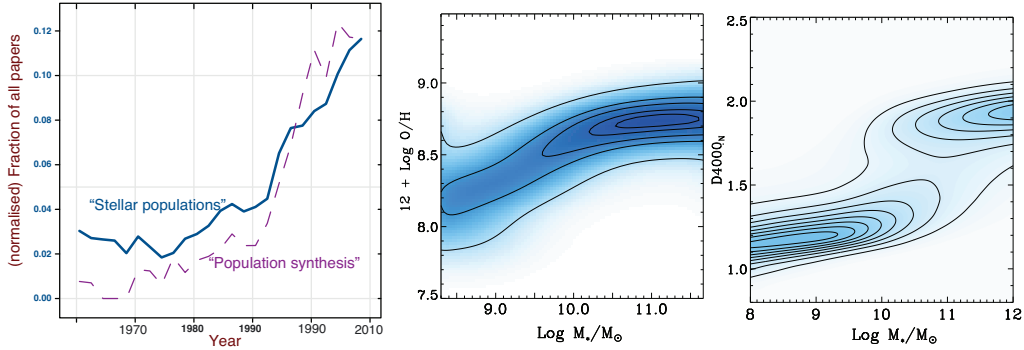} 
 \caption{\textit{Left panel:} The fraction of papers in A\&A, AJ, ApJ and MNRAS each year
   that mention ``Stellar populations'' (solid line) or ``Population
   synthesis'' (dashed line, scaled by a factor of 8) in their
   abstracts. Currently about 12\% of all papers mention ``Stellar
   populations'' and 1.5\% ``Population synthesis''.  
   \textit{Middle panel:}  The mass-metallicity relation for SDSS DR6 (cf.\ Tremonti et al
    2004 for DR2) with a metallicity adjustment following 
    \citet{2008A&A...485..657B}. \textit{Right panel:} The relation between stellar mass and the 4000\AA\   break in the SDSS DR7. This is an updated version of the
    influential study by~\citet{2003MNRAS.341...54K}, showing the bi-modality of the galaxy population.} 
   \label{fig1}
\end{center}
\end{figure}

\subsection{The estimation of stellar masses and the mass assembly of
  the Universe} 
\label{sec:estim-stell-mass}

%

There is now a vast body of deep photometric observations of the
sky\footnote{For a compilation see
  \texttt{http://www.strw.leidenuniv.nl/{\urltilda}jarle/DeepFields/}} covering
a large stretch of cosmic time. By matching these observations to
stellar population models, it is possible to infer the stellar mass of
the distant objects
\citep[e.g][]{1998AJ....115.2169G,2000ApJ...536L..77B,2001ApJ...550..212B,2001ApJ...562L.111D,2003ApJ...594L...9F,2003MNRAS.341...33K,2003ApJ...587...25D,2006ApJ...652...85M}. This
has led to stellar mass becoming the most important independent
variable for galaxy evolution studies. Notable applications of this is
the quantification of a transition mass in the local galaxy population
\citep{2003MNRAS.341...54K} and the mass-metallicity relation
\citep{2004ApJ...613..898T}, updated versions of the key plots from
those studies is shown in Figure~\ref{fig1}.

When the stellar masses of galaxies have been calculated for a
statistically well-defined sample, it is possible to calculate the
stellar mass density in galaxies. By combining samples over a wide
range in redshift, we can infer the stellar mass assembly history of
the Universe \citep[e.g.][see Wilkins et al 2008 for a
compilation]{2000ApJ...536L..77B,2003ApJ...587...25D,2003ApJ...594L...9F,2007A&A...476..137A}\nocite{2008MNRAS.385..687W}. 

This has provided a very visual image of the assembly of mass, and
with the large samples currently available it is possible to do this
in bins of mass and hence derive the evolution of the stellar mass
function with cosmic time \citep[e.g.][]{2009ApJ...701.1765M} and we
now know that the massive end of the stellar mass function was in
place at much higher redshift than was expected a decade ago.

\section{The future --- context}
\label{sec:future-context}

Major progress has come in extra-galactic research in the recent
decade through large-scale surveys of the sky, such as the 2dF Galaxy Redshift Survey
\citep{2001MNRAS.328.1039C} and the Sloan Digital Sky Survey
\citep[SDSS,][]{2000AJ....120.1579Y} in the nearby Universe, and the
Vimos Very Deep Survey \citep[VVDS, ][]{2004A&A...428.1043L} and
DEEP-2 survey \citep{2003SPIE.4834..161D} out to $z\sim 1$. The coming
years will see a continuation of this trend with, for instance, the
commencement of the various VISTA and VST surveys
\citep[see][]{2007Msngr.127...28A}, and several Dark Energy surveys
(e.g. BOSS\footnote{\texttt{http://cosmology.lbl.gov/BOSS/}},
WiggleZ\footnote{\texttt{http://wigglez.swin.edu.au/Welcome.html}} and
HETDEX\footnote{\texttt{http://www.as.utexas.edu/hetdex/}}). These developments
are very exciting and it is clear that stellar population modelling
will be key to the exploitation of the data coming out of these
studies. 

However, while these large-scale studies undoubtedly will be very
valuable for a wide range of scientific questions, they do not present
stellar populations with very different challenges from what we
already face and hence I will not focus on these surveys here. The situation is somewhat different with the Large Synoptic Survey Telescope\footnote{\texttt{http://www.lsst.org/lsst}} (LSST), which, if built, will provide repeat imaging of the northern sky --- opening up the possibility for a systematic use of variability in the study of stellar populations. This is clearly a challenge for the coming decade but not one I will discuss further here. Neither will I discuss the opportunities offered by new long-wavelength facilities such as Herschel, ALMA or SKA, or high energy facilities such as Fermi and IXO, even though these all offer novel challenges and in the future we will hopefully see them integrated into stellar population analysis \citep[c.f.][]{2008MNRAS.388.1595D}.


Instead I will focus on the major change in studies of the
extra-galactic Universe that will be ushered in with the launch of
the James Webb Space Telescope \cite[JWST, 
][]{2006SSRv..123..485G} and the building of one or more Extremely
Large Telescopes (ELTs) such as the European ELT\footnote{\texttt{http://www.eso.org/sci/facilities/eelt/}}, the TMT\footnote{\texttt{http://www.tmt.org}} and the GMT\footnote{\texttt{http://www.gmto.org}}.  These facilities will present a significant shift in emphasis relative to much of the stellar population modelling carried out at present.

The JWST will be a truly transformative influence on astronomy and will
firmly shift the focus of extra-galactic observations to longer wavelengths than is commonly used today. At wavelengths $\lambda > 2\mu$m JWST will be two orders of magnitude more sensitive than current facilities, opening up an entirely new area of wavelength space for extra-galactic studies.


The proposed ELTs will have two advantages relative to today's 10m-class telescopes: Much higher sensitivity and much higher spatial resolution. The former will enable us to obtain much larger samples of spectra of distant galaxies and detect much fainter and more distant galaxies than we currently can. The improved resolution will be achieved with the help of adaptive optics systems working in the near- and mid-IR, and this will enable resolved stellar population studies to reconstruct star formation histories for more distant galaxies and in more crowded regions. 

Thus, while extra-galactic research in the coming decade will make use of much larger samples and the routine observations of very faint objects, the main gain relative to today will come from studies in the near- and mid-IR; both because JWST will operate at those wavelengths, but also because those are the wavelengths where adaptive optics, and hence improved spatial resolution, will work best. The challenge for stellar population studies will be to exploit these data in an optimal manner.  I should emphasise that this does not mean that optical astronomy will become obsolete, merely that I don't view that wavelength range as presenting as many \emph{new} challenges as  longer wavelengths and hence of less importance for this review, although it will be discussed below.

\section{Challenges --- or areas of interest}
\label{sec:challenges-or-areas}

\subsection{Error estimates on stellar population models}
\label{sec:error-estim-stell}

Models for stellar populations have reached a high degree of
sophistication and ease of use. As mentioned above, this has led many researchers to use them to help interpret their observational data. With this wide-spread
adoption comes an increased need to quantify the \emph{uncertainties}
in the models as well --- stellar population models are rather
involved and it is difficult for a user to assess the
reliability of a particular prediction without input from the model
builders. These error sources are many: Uncertainties in the stellar evolution models for normal stars, such as the treatment of turbulent mixing; uncertainties in atomic data; treatment of uncertain or intractable phases of stellar evolution such as horizontal branch (HB) stars, thermally pulsing asymptotic giant branch (TP-AGB) stars, post-AGB stars or various phases of binary evolution; stellar wind loss; uncertainties in stellar atmosphere calculations; mismatches between stars in observational libraries and the theoretical tracks they are matched to and observational uncertainties in the empirical data included in the models --- just to mention a few.

These \textit{model} uncertainties are likely to dominate the error budget in analyses of high S/N galaxy spectra, and until we understand them, we will not be able to take full advantage of the best data a telescope can deliver. Thus a major challenge for model builders in the coming decade is to
construct models that incorporate uncertainty estimates, or probably
more realistically, indications of reliability for different
predictions. Notable first attempts at this has been recently
published by \citep{Conroy2009a}, who explored the consequences of uncertainties in out understanding of HB and TP-AGB stars on the predictions of population synthesis models, and Percival \& Salaris (2009), who explored the impact of uncertainties in the calibration of fundamental stellar parameters. These are promising first steps but are still far from a comprehensive study. This is likely to improve in the future but an accurate treatment of uncertainties is likely to remain a major challenge for years to come.

\subsection{The near-IR --- can we properly make use of future observations?}
\label{sec:near-ir-can}

We emphasised  above that adaptive optics on ELTs and in
particular the flight of JWST will ensure that much of the focus for future
astronomical observations will move into the near- and mid-IR, 
both for photometry and spectroscopy. It is therefore
essential to ensure that models accurately reproduce the rest-frame
near- to mid-IR fluxes of stellar populations, and to understand
better what near-IR spectral features are of importance for studying
stellar populations.

We expect that resolved stellar population studies making use of the much improved resolution offered by adaptive optics systems on ELTs will be widespread in the future. It is therefore paramount that we start carrying out similar studies with current facilities to learn how to make optimal use of ELTs for this kind of work. The results from such efforts have
only recently started to appear. An example of this is the study of
Galactic globular clusters by \citet[][NGC6440]{2008ApJ...687L..79O}
and \citet[][NGC 6388]{2009A&A...493..539M} using NACO on the VLT, and the studies by
Fiorentino et al (in prep.) of NGC 1928 and \citet{2008SPIE.7015E..61C}
of R136 using the Multi-Conjugate Adaptive Optics Demonstrator \citep[MAD,][]{2007Msngr.129....8M} on the VLT. While very useful, these latter two studies
in the Magellanic clouds do not yet probe down to the main-sequence
turn-off and hence do not put very strong constraints on models, hopefully this can be remedied in the near future with deeper observations.

Resolved stellar population studies in nearby galaxies will not only
be carried out using photometric observations, but also using adaptive
optics allied to near-IR IFUs, such as the proposed EAGLE for the
E-ELT, IRS/IRMOS on the TMT and GMTIFS on the GMT. To make optimal use
of these facilities it is clearly necessary to  make use of spectral features in the near-IR. This is an area that has seen
significant effort recently, with in-depth studies of the CO band-head
at 2.3$\mu$m by e.g. \citet{Marmol-Queralto2008}, and assembly of
near-IR atlases of stellar spectra such as the IRTF Spectral Library
\citep{2009arXiv0909.0818R}. At the moment these studies cover
relatively restricted ranges in stellar parameter space and we need a considerable effort both on the theoretical and observational side in the coming
decade to ensure that we can make optimal use of cutting-edge
facilities on ELTs.

Studies of resolved stellar populations can often ignore problematic stellar phases by excluding them from analysis. For the majority of galaxies, which are unresolved, there is no such option. Thus fitting unresolved stellar populations in the near-IR can be expected to be challenging. This is borne out by experience, although for old stellar populations (ellipticals/S0s), \citet{2009MNRAS.397..695C} showed that a wide range of population synthesis models give consistent results when applied to broad-band photometry including near-IR.  In general, the situation is more complex as was shown, for instance, by \citet{2008MNRAS.384..930E} in their comparison between models and UKIDSS + SDSS photometry. They pointed out that while some combinations of near-IR and optical filters could be well-fit by models, others could not.

%

To focus our discussion further, let us consider the determination of stellar masses. As emphasised above, this is an essential ingredient of most galaxy evolution studies today and it is therefore important to understand what systematic uncertainties can influence their determinations. The estimation of stellar masses fundamentally involve a fit to either broad-band photometry or spectral features to estimate $M_*/L_X$ in some band, $X$, and finally this is scaled by $L_X$ to give the stellar mass. Any systematic uncertainties in the models might lead to biases in the estimates of $M_*/L_X$. 

The most obvious systematic uncertainty is probably the IMF, but we will ignore this here due to space limitations. Focusing instead on the population synthesis models, \citet{2009ApJ...699..486C} showed that systematic uncertainties in stellar population models probably limit the accuracy of stellar mass determinations to about 0.3 dex. In addition to this we know that  different algorithms for how star formation histories are treated can lead to systematic differences of the order of 0.2 dex \citep{2007A&A...474..443P}, but that when the algorithms are the same, different models give the same stellar mass to within 0.15 dex (based on a comparison of the mass estimates from Bruzual \& Charlot 2003 models (BC03) and Maraston 2005 models (M05) by  Tojeiro et al 2009) \nocite{2009arXiv0904.1001T} \nocite{2003MNRAS.344.1000B}\nocite{2005MNRAS.362..799M}.

Many studies looking at the impact of including rest-frame near-IR data in the fit have focused on higher redshift galaxies. They generally find that the situation is less good. \citet{2006ApJ...652...97V} compared dynamical and stellar mass estimates at both low and high redshift and found that models did not give acceptable results when near-IR data was included. More recently, e.g.\ \citet{2009ApJ...701.1839M} found that model predictions in the near-IR were the limiting systematic uncertainty in their study of $z\sim 2.3$ galaxies. Some of this could be due to the way very young stars are treated in the  models as  Conroy et al (2009) showed that there is a glaring discrepancy between the observed $V-K$ colour of LMC clusters and the predictions of models at young ages, but we note that  this relies age measurements from a broad range of sources with rather different age estimation techniques and dust corrections are non-trivial. A careful re-examination of this issue might be worthwhile.  However it is generally accepted that part of this discrepancy in stellar mass estimates is likely to be due to the treatment of TP-AGB stars in the models. These stars dominate the near-IR luminosity of stellar populations with an age of a few Gyr. Since this is the typical age of galaxies at $z\sim 2$--$3$, it follows that uncertainties in the treatment of this population in the models can have severe impact on the interpretation of observational data. This was emphasised by \citet{2006ApJ...652...85M} who highlighted the differences in the treatment of TP-AGB stars in M05 and BC03. Subsequent work has done much to improve the situation and Marigo \& Girardi have recently published improved evolutionary tracks including a significantly improved treatment of TP-AGB stars  \citep[][see also the contribution by Margio in these proceedings]{2008A&A...482..883M,2007A&A...469..239M}. Despite this, it is fair to say that there is still considerable uncertainty associated to the treatment of TP-AGB stars at low metallicity and high redshift. It is difficult to see that major progress can be made with resolved stellar populations alone, rather one might have to combine resolved stellar population studies with ''almost resolved'' studies using surface brightness fluctuation studies in the near-IR and optical \citep[e.g.][]{2009ApJ...700.1247R,2009arXiv0902.1177L}.

\subsection{Non-solar abundance ratios --- interpreting high S/N data}
\label{sec:non-solar-abundance}

While most population synthesis models assume scaled-solar abundance
ratios for their stellar tracks and atmospheres, it has long been
recognised that variations in the abundance ratios of 
elements away from solar ratios can strongly affect any inferences one
makes based on spectral features.  Deviations from scaled-solar
abundance ratios reflect a variation in past star formation histories, in particular the relative
contributions by Supernova Type II and Ia
\cite[e.g.][]{1992ApJ...398...69W,Trager2000,Thomas2003}.  This is normally seen where the typical time-scale for star formation is shorter than $\sim 1$Gyr \citep[e.g.][]{2001ApJ...558..351M}, as in elliptical galaxies \citep{Thomas2005} or at $z>4$.

Treating non-solar abundance variations has long been a challenge for
stellar population models, but the last decade, in particular, has seen
great progress. There is now a fairly well-developed machinery for
dealing with $\alpha$-variation within the Lick index system using
fitting functions
\citep[e.g.][]{1995AJ....110.3035T,Trager2000,Thomas2003,2007A&A...462..481T,2007ApJS..171..146S}
and this has been extensively used to study the properties of massive
galaxies in the nearby Universe
\citep[e.g.][]{Thomas2005,2009ApJ...693..486G,2009ApJ...698.1590G,2009arXiv0908.2990S}.

There has also been great progress in the calculation of theoretical
isochrones for non-solar abundance ratios
\citep{2004ApJ...612..168P,2004ApJ...616..498C,2006ApJ...642..797P,2007IAUS..241...43W,2007ApJ...666..403D}
and there has also been significant progress in the calculation of
stellar atmosphere models
\citep[e.g.][]{2005A&A...443..735C,2005A&A...442.1127M}, see
\citet{2007MNRAS.381.1329M} for an in-depth comparison of some of
these libraries.  Taken together this has allowed the construction of libraries suited
to population synthesis of integrated populations
\citep[e.g.][]{2007MNRAS.382..498C,2009ApJ...690..427P}. These models are now
sufficiently mature that they can begin to be used for the
interpretation of galaxy spectra \citep[e.g.][]{Walcher2009}.

Despite the great progress made, there are still some major issues
outstanding. The first is that considerable work must be done to
understand what detailed features in the theoretical libraries can be
fully trusted --- this needs careful testing against high-resolution
spectroscopy. Related to this, and also of great importance and
interest, is the question of whether it is sufficient to just vary the
$\alpha$-elements as a block, or whether the individual elements must
be varied one by one. It has been argued that the latter is indeed necessary
to optimally extract information from high-S/N spectra, and some
progress has been made recently
\citep{2007ApJ...666..403D,2009ApJ...694..902L}. This does expand the
available parameter space significantly and should be applied in a
careful manner.

\subsection{The first stars and galaxies}
\label{sec:first-stars-galaxies}

One of the major challenges for stellar populations in general will be
to interpret observations of the very high-redshift Universe from
JWST. It is clear that it will become necessary to understand what
kind of objects we are likely to see, will they be truly zero
metallicity, or will they have very small metallicities? And even
within the zero metallicity class there is now thought to be two
distinct classes of stars, referred to as Pop III.1 and Pop III.2 stars
by \citet*{2008AIPC..990...47T}, depending on whether or not their formation
was affected by radiation by an earlier population 
\citep{2008ApJ...681..771M,2006MNRAS.373..128G}.  This in turn has
important implications for the IMF of the first
stars, and therefore on the strength of the He\textsc{ii} 1640\AA\
line which has often been suggested as a probe of Population III stars
\citep{2001ApJ...550L...1T,2003ApJ...584..608T,2003A&A...397..527S}. 

In addition to these uncertainties, it has also recently been realised
that the halos where the first stars form have a very high dark matter
density. If the dark matter is its own anti-particle,
self-annihilation is a possible energy source for the first
stars, a topic that has seen considerable interest recently \citep[e.g.][]{Spolyar2008,Freese2008,2008MNRAS.390.1655I,Ripamonti2009}. 
It is not clear whether the effect of dark matter annihilation on the first stars will
turn out to be crucial, or indeed observable. However it has been argued that similar effects might be observable  in the Galactic centre \citep{Scott2009,Fairbairn2008}, thus this is an area with potential future implications.

\subsection{The rest-frame ultraviolet --- understanding massive stars}
\label{sec:rest-UV}


In any galaxy with on-going star formatioan, the rest-frame ultraviolet (rest-UV) spectrum is dominated by O and B stars. This can be viewed as both a disadvantage ---- our understanding of massive stars, whether their evolution is strongly influenced by rotation (e.g. Meynet et al 2009), or significantly influenced by binary evolution (e.g. Eldridge et al 2008), is certainly lacking in some aspects. On the other hand it is an advantage because it offers us a direct probe of massive stars, providing us with a wealth of information to test and develop models. With the upcoming demise of HST, the coming decade will see a dearth of UV-sensitive facilities, but optical spectrographs on ELTs will provide very large samples of optical spectra of galaxies at $z\sim 2$--$3$, sampling the rest-UV.  

Thus it is a major challenge for stellar population models to be able
to extract information from these spectra (e.g. Rix et al 2004; Maraston et al 2009). As discussed by Leitherer (these proceedings), stellar atmosphere models for hot, massive stars are now reaching maturity and will likely be sufficient for most future studies. On the other hand, the evolution of massive stars, their mass-loss prescriptions and disentangling the influence of rotation and of binaries remain a challenge.

It is interesting in this context to note that the light-gathering power of a 42m-class telescope is about 25 times larger than that of an 8m-class telescope (for resolved
sources). This is similar to the amplification seen in
gravitationally lensed Lyman-break galaxies \citep{2007ApJ...662L..51A,2007ApJ...671L...9B,2007ApJ...654L..33S,2009MNRAS.392..104B,2009ApJ...699.1242L,2009ApJ...696L..61K}.  Thus by observing these objects, it is possible to test models on very similar data to
that which will be obtained routinely with an ELT, something that several groups have realised \cite[e.g.][]{2000ApJ...528...96P,2009MNRAS.398.1263Q,2009ApJ...701...52H}. This should clearly continue and a concerted effort for a survey of the nearby Universe with COS would also be very useful.

\subsection{The importance of binaries}
\label{sec:binaries}

Most stars with mass $>1\Msun$ are born in binaries and the binary
fraction appears to increase with stellar mass
\citep[e.g.][]{2006ApJ...640L..63L}, thus there is no doubt that
binary evolution must be important for understanding the stellar
populations of galaxies at some level (see the contribution by
Vanbeveren in these proceedings). However the inclusion of binaries
into population synthesis models is problematic as it leads
to a much increased parameter space and complexity.

Including binary evolution is however crucial to understand some rare, but bright, evolutionary phases. This includes certain variable star phases, X-ray binaries and stellar remnants, most of which should ideally be included in population synthesis modelling. However in the UV-optical region, the main concern is with horizontal branch stars  (see Moni Bidin et al 2008 for a review). These stars dominate the near-UV light in old stellar populations and understanding whether they are due to binary evolution (e.g. Han et al 2007)or not has important implications for the analysis of elliptical galaxies.

Binary evolution might also be crucial for the development of Wolf-Rayet stars, particularly at low metallicities. Brinchmann et al (2008) showed that large number of Wolf-Rayet stars at low metallicity is a significant challenge to stellar evolution models.  Eldridge \& Stanway (2009) recently showed that this tension can be resolved by binary evolution effects, although similar results can be achieved by inclusion of rotation in the evolution of massive stars (e.g. Meyner \& Maeder 2005). The relative importance of these two effects is not yet understood and is an important challenge for the future.

\section{Conclusions}
\label{sec:conclusions}

Rather than a traditional conclusion, I would here like to end by summarising the challenges above into a set of rest wavelength ranges, highlighting the issues and some suggestions as to what might be useful studies to do --- note that most of these suggestions are merely reinforcing already existing studies!

\textbf{Far-UV}  ---  \textit{Massive stars --- what are realistic rotation and binary parameters}. \textbf{To do:}  Observations of lensed Lyman-break galaxies, COS observations of nearby stars and star-forming regions,  more in-depth studies of stellar mass-loss 

 \textbf{Near-UV}  ---  \textit{Binary evolution, mass loss on the RGB}. \textbf{To do:}  COS and GALEX spectroscopy of UV-upturn galaxies, inclusion of horizontal branch uncertainties in models. Combination of optical, near-UV and X-ray data in analysis.

\textbf{Optical} --- \textit{High-precision predictions. Non-solar abundance variations and impact on spectra,  emission lines}. \textbf{To do:} Careful and extensive testing of theoretical spectra, comparison to detailed spectroscopic analysis of nearby systems: what abundances can be reliably extracted from medium resolution spectra?

\textbf{Near-IR}  ---  \textit{Evolution of and spectra of luminous, cold stars. TP-AGB stars, RGs and RSGs, spectral features in the near-IR}. \textbf{To do:} Further observations of resolved stellar populations in the near-IR with and without adaptive optics with careful comparison to results including optical data. Calibration spectral features in the near-IR. Obtain ACS optical data to back up future studies in near-IR with ELTs. In depth analysis of uncertainties in population synthesis models and careful comparisons to data in the nearby and distant Universe.

\begin{multicols}{2}

\end{multicols}


\end{document}